\documentclass[10pt,a4paper,conference]{IEEEtran}

\usepackage{amsfonts,amssymb}
\usepackage{latexsym}
\usepackage{amsmath}
\usepackage{graphicx}   
\usepackage{mathrsfs}  
\usepackage{psfrag} 

\newcommand{\stxt}[1]{\ensuremath{_{\text{#1}}}}
\newcommand{\vect}[1]{\mbox{\boldmath $#1$}}
\newcommand{\Tr}{\ensuremath{^{\mathrm{T}}}}
\newtheorem{theorem}{Theorem}
\newtheorem{lemma}{Lemma}
\newtheorem{prop}{Proposition}
\newcommand{\beq}{\begin{equation}}
\newcommand{\eeq}{\end{equation}}
\newcommand{\beqa}{\begin{eqnarray}}
\newcommand{\eeqa}{\end{eqnarray}}

\begin{document}

\title{Terminated LDPC Convolutional Codes with Thresholds Close to Capacity}

\author{\authorblockN{\hspace{-5ex}Michael Lentmaier\authorrefmark{1}, Arvind Sridharan\authorrefmark{2}, Kamil Sh.~Zigangirov\authorrefmark{3}, and Daniel J.~Costello, Jr.\authorrefmark{3}}
\authorblockA{\authorrefmark{1}German Aerospace Center (DLR)$^1$,
 Inst. for Communications and Navigation, Email: {\tt Michael.Lentmaier@dlr.de}}
\authorblockA{\authorrefmark{2}Seagate Technology, 389 Disc Drive,
Longmont, CO 80503, USA,
Email: {\tt Sridharan.1@nd.edu}}
\authorblockA{\authorrefmark{3}Dept. of Electrical Engineering, University of Notre Dame, USA, Email: {\tt \{Zigangirov.1,Costello.2\}@nd.edu}}
}
%

\maketitle
\footnotetext[1]{Michael Lentmaier was with the Dept. of Electrical Engineering at University of Notre Dame, USA and with the Dept. of TAIT at University of Ulm, Germany.}
\begin{abstract}
An ensemble of LDPC convolutional codes with parity-check matrices composed of permutation matrices is considered. The convergence of the iterative belief propagation based decoder for terminated convolutional codes in the ensemble is analyzed for binary-input output-symmetric memoryless channels using density evolution techniques. 
We observe that the structured irregularity in the Tanner graph of the codes leads to 
significantly better thresholds when compared to corresponding LDPC block codes.
\end{abstract}

\section{Introduction}
Low-density parity-check (LDPC) block codes, invented by Gallager~\cite{gallager}, have been shown to achieve excellent performance on a wide class of channels. 
The convolutional counterparts of LDPC block codes, LDPC convolutional codes, have been described in~\cite{fels_zig}\cite{ldcc_1}\cite{alg_LDPC}. Both LDPC block and convolutional codes are defined by sparse parity-check matrices and can be  decoded iteratively using message passing algorithms (e.g., belief propagation) with complexity per bit per iteration independent of the block length or constraint length. This makes iterative decoding of LDPC codes with large block length or constraint length feasible.

In~\cite{ldcc_2}, the existence of a sequence of $(J,K)$ regular\footnote[2]{$(J,K)$ regular LDPC codes are defined by parity-check matrices having  $J$ ones in each column of the matrix and $K$ ones in each row of the matrix.} 
LDPC convolutional codes for which an arbitrary number of independent iterations is possible was demonstrated. Based on this result, it follows that the threshold of $(J,K)$ regular  LDPC convolutional codes is at least as good  
as the threshold of $(J,K)$ regular LDPC block codes for any message passing algorithm and channel.
 Moreover, simulation results on the additive white Gaussian noise channel (see~\cite{ldcc_1}\cite{alg_LDPC}) indicate the possibility that LDPC convolutional codes may have better thresholds than corresponding LDPC block codes. 

In this paper we consider a class of regular LDPC convolutional codes with parity-check matrices 
composed of blocks of randomly constructed $M \times M$ permutation matrices. For the erasure channel, iterative belief propagation  decoding of terminated LDPC convolutional codes
in this class was analyzed in \cite{allerton}. 
There it was shown that the termination leads to a structured irregularity in the Tanner graph, 
and that this structured irregularity 
leads to significantly better thresholds compared to
corresponding randomly constructed regular and irregular LDPC block codes. Further, it was observed that 
the thresholds approach the capacity of the erasure channel. 
In this paper we generalize the techniques of  \cite{allerton} to 
arbitrary binary-input memoryless channels and give numerical examples for the AWGN channel. 

\section{Convolutional Code Ensemble}
A rate $R=b/c$ binary convolutional code can be defined as the set of sequences $\vect{v}=(\dots,\vect{v}_{-1},\vect{v}_0,\vect{v}_1,\dots), \ \vect{v}_t \in \mathbb{F}_2^{c}$, 
satisfying the equality $\vect{vH}\Tr=\vect{0}$, where the infinite syndrome former matrix $\vect{H}\Tr$ is given by
\[
\vect{H}\Tr=\begin{pmatrix}
\ddots & & \ddots & & \\
\vect{H}_0\Tr(0) & \dots & \vect{H}_{m\stxt{s}}\Tr(m\stxt{s}) & &\\
 &\ddots & & \ddots & \\
 && \vect{H}_0\Tr(t) & \dots & \vect{H}_{m\stxt{s}}\Tr(t+m\stxt{s})\\
 && \ddots & & \ddots \\
\end{pmatrix}\ ,
\label{eqn:H}
\]
and each $\vect{H}_{i}\Tr(t+i)$ is a $c \times (c-b)$ binary matrix. If $\vect{H}\Tr$ defines a rate $R=b/c$ convolutional code, the matrix $\vect{H}_0\Tr(t)$ must have full rank for all time instants $t$. In this case, by suitable row permutations we can ensure that the last $(c-b)$ rows are linearly independent. Then the first $b$ symbols at each time instant are information symbols and the last $(c-b)$ symbols the corresponding parity symbols. The largest $i$ such that $\vect{H}_{i}\Tr(t+i)$ is a nonzero matrix for some $t$ is called the syndrome former memory $m\stxt{s}$.

\begin{figure}\centering
\scalebox{0.7}{
\begin{psfrags}
\psfrag{H =}[c][r]{\hspace*{-10ex}$\vect{H}\Tr=$}
\psfrag{MxM permutation}[c][c]{\rule[-1em]{0mm}{0mm}$M \times M$ permutation}
\psfrag{matrices}[c][c]{\hspace*{2ex} matrices}
\psfrag{J=3}[l][c]{\hspace*{-5ex} $J=3$} \psfrag{K=6}[r][c]{$K=6$}
\includegraphics[width=0.5\linewidth]{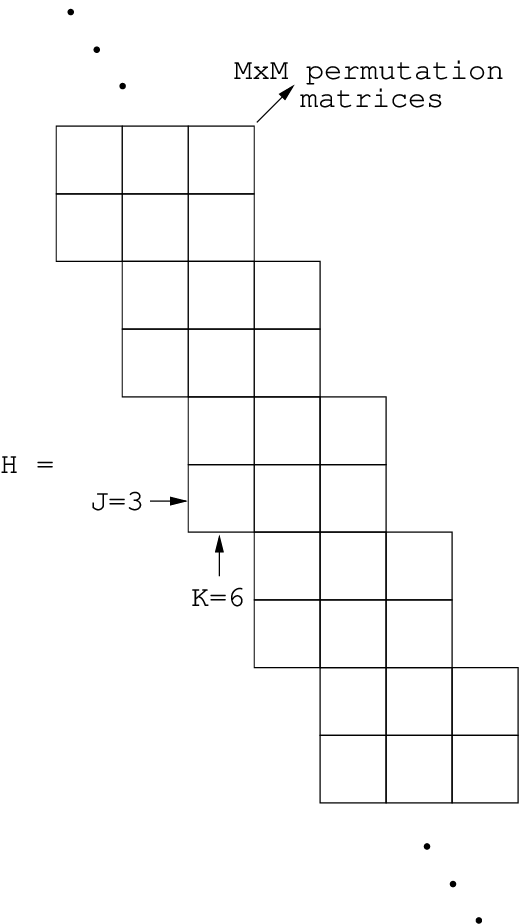}
\end{psfrags}}
\caption{\small{Syndrome former matrix for a code in $C_P(3,6,M)$.}}
\label{fig:conv_syn}
\vspace*{-1.5em}
\end{figure}
LDPC convolutional codes have sparse syndrome former matrices. A $(J,K)$ regular LDPC convolutional code is defined by a syndrome former that contains exactly $J$ ones in each row and $K$ ones in each column.

We now define the ensemble of LDPC convolutional codes of interest. Though the ensemble can be defined more generally, in this paper we focus on the case $K=2J$, $J > 2$. 
We construct LDPC convolutional codes  defined by syndrome formers $\vect{H}\Tr$ with syndrome former memory $m\stxt{s}=J-1$. 
For $i=0,1,\dots,J-1$, the sub-matrices $\vect{H}_{i}\Tr(t+i)$ of the syndrome former are the matrices $\left(\vect P^{(0)}_{i}(t+i), \vect P^{(1)}_{i}(t+i)\right)\Tr$, where $\vect P^{(h)}_{i}(t+i),\ h=0,1,$ is an $M \times M$ permutation matrix. All other entries of the syndrome former are zero matrices. 
 Equivalently, 
each $\vect{H}_{i}\Tr(t+i)$, $i=0,1,\dots,J-1$, is a $c \times (c-b)$ binary matrix, where $c=2M$ and $b=M$. 
By construction it follows that each row of the syndrome former $\vect{H}\Tr$ has $J$ ones and each column $K$ ones. Let $\mathcal {C}_P(J,2J,M)$ denote this ensemble of $(J,2J)$ regular LDPC convolutional codes. (Note that the ensemble of codes $\mathcal {C}_P(J,2J,M)$ is time-varying.) 
Fig.~\ref{fig:conv_syn} shows the syndrome former matrix of a $(3,6)$ regular LDPC convolutional code in $\mathcal {C}_P(3,6,M)$.

Since $\vect{H}_{0}\Tr(t)$ consists of two non-overlapping permutation matrices, it has full rank. Hence $\vect{H}\Tr$ defines a rate $R=\frac{M}{2M}$ code. 
Further, the constraint imposed by the syndrome former, i.e.,
\beq
{\vect{v}_t\vect{H}_0\Tr(t)+\vect{v}_{t-1}\vect{H}_1\Tr(t)+\dots+\vect{v}_{t-m\stxt{s}}\vect{H}_{m\stxt{s}}\Tr(t)=\vect{0}, }
\label{eqn:sys_enc}
\eeq 
where $\vect{v}_t \in \mathbb{F}_2^{2M}, t \in \mathbb{Z}$,
can be used to perform a systematic encoding of the code~\cite{fels_zig}. 
The constraint length of codes in $\mathcal {C}_P(J,2J,M)$ is defined as $\nu=(m\stxt{s}+1) \cdot c=J \cdot 2M=KM$. Thus, the constraint length of codes in the ensemble $\mathcal {C}_P(3,6,M)$ is 
$6M$. 

The Tanner graph for a code in $\mathcal {C}_P(J,2J,M)$ can be obtained from its syndrome former matrix. The graph consists of symbol and check nodes, each symbol node corresponding to a particular row and each check node corresponding to a particular column of the syndrome former matrix $\vect H\Tr$. 
There is an edge between a symbol node and a check node if the corresponding symbol takes part in the respective parity-check equation. 
For the Tanner graph of a convolutional code we can associate a notion of time. At each time instant $t$ the sub-matrices $\vect H_i(t)$ of the syndrome former $\vect H\Tr$  lead to $c-b$ check nodes in the Tanner graph. Similarly, for each time instant $t$, 
there are $c$ symbol nodes in the Tanner graph. Observe that $\vect H_i(t)$ is non-zero only from $i=0,1,\dots,m\stxt{s}$, and hence nodes in the Tanner graph can be connected at most $m\stxt{s}$ time units away. The Tanner graph of a code in the ensemble $C(M,2M,3)$ is comprised of $c=2M$ symbol nodes and $c-b=M$ check nodes for each time instant. Further, each node can be connected at most $m\stxt{s}=2$ time units away.

For practical applications, a convolutional encoder starts from a known state (usually the all-zero state) and, after the data to be transmitted has been encoded, the encoder is terminated back to the all-zero state. It can be shown that for the ensemble $\mathcal {C}_P(J,2J,M)$ we need a tail for no more than $m\stxt{s}+1$ time instants, i.e., $(m_s+1) M$ information bits to return the encoder to the all-zero state~\cite{thesis}.

\begin{figure}
\scalebox{0.7}{
\begin{psfrags}
\psfrag{time }{time}
\psfrag{2M}[c][c]{$2M$}
\psfrag{L}[c][c]{$L+3$}\psfrag{t=1}[c][c]{$t=1$}\psfrag{t=2}[c][c]{$t=2$}\psfrag{t=L+2}[c][c]{$t=L+2$}\psfrag{t=L+3}[c][c]{$t=L+3$}
{\includegraphics[width=1.4\linewidth]{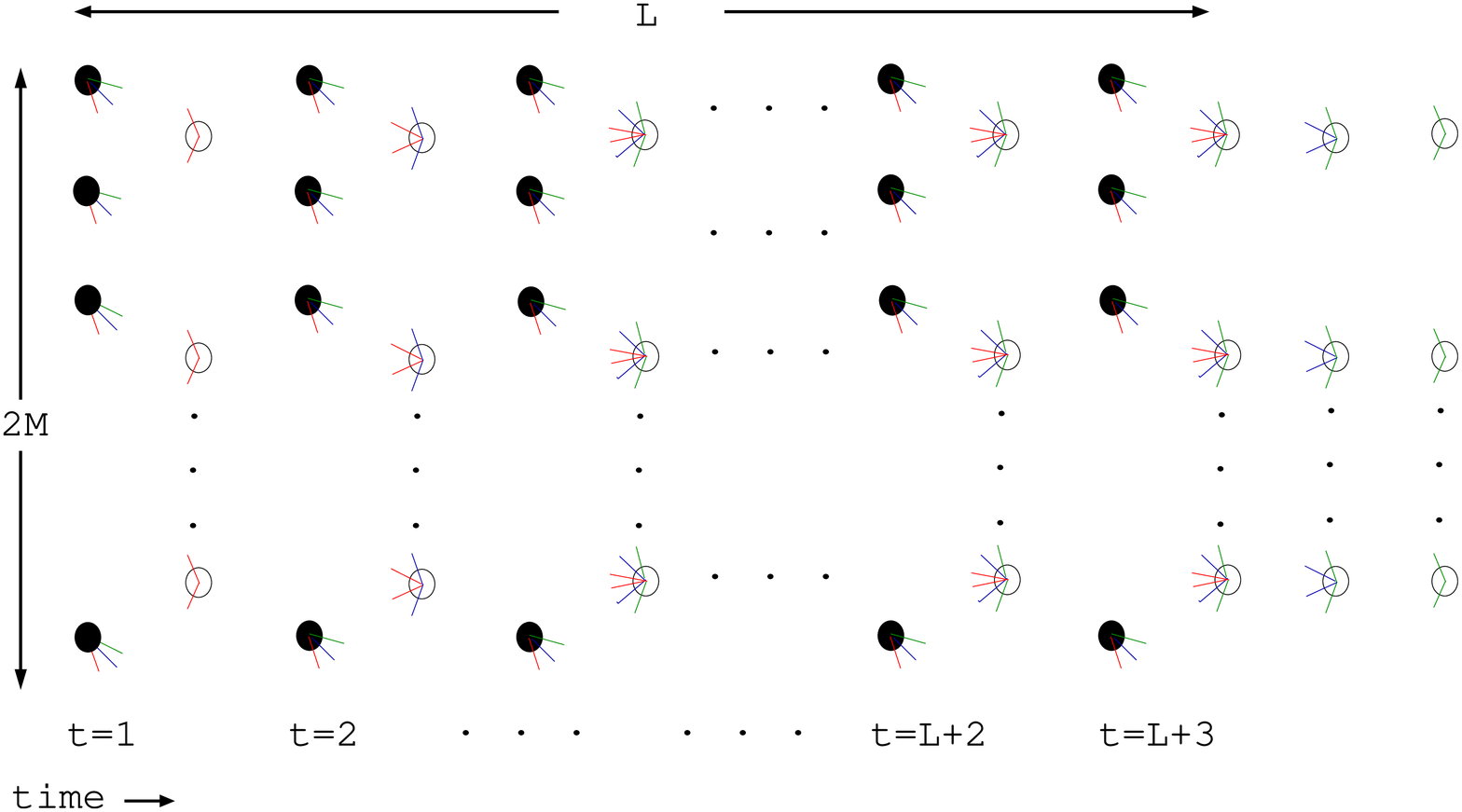}}
\end{psfrags}}
\caption{Tanner graph of a terminated convolutional code obtained from $\mathcal {C}_P(3,6,M)$.}
\label{fig:term_tann_graph}
\vspace*{-1.5em}
\end{figure}
Suppose that we wish to transmit $LM$ information bits using a code from $\mathcal {C}_P(J,2J,M)$. It follows that the terminated code has rate $R=0.5/(1+\frac{J}{L})$. Note that for $L >> J$, the rate loss is negligible. In Fig.~\ref{fig:term_tann_graph} we show the Tanner graph of a terminated code obtained from a convolutional code in the ensemble ${C}(M,2M,3)$. 
Observe that symbols are zero both before encoding begins, i.e., $t=1$, and after termination, i.e., $t=L+3$. Hence in obtaining the Tanner graph of the terminated convolutional code, edges connecting check nodes to any of the symbol nodes that are known to be zero can be omitted. For example, we can disconnect the check nodes at time $t=1$ from symbol nodes at time $t<1$, since these are known to be zero. It follows that, while all symbol nodes in Fig. \ref{fig:term_tann_graph} have degree three, the check nodes can have degree either two, four, or six. Note that, even though the convolutional code is regular, knowing bits perfectly before encoding and after termination leads to a slight irregularity in the Tanner graph of the terminated convolutional code.

\section{Decoding Analysis for Binary-Input Memoryless Channels}
As for block codes, an iterative decoder for LDPC convolutional codes can be conveniently  described on the Tanner graph.
In each decoding iteration messages are exchanged between the symbol nodes and the check nodes.
We consider an algorithm equivalent to the probabilistic iterative decoding algorithm proposed by Gallager, which in a wider context is known as belief propagation or the sum-product algorithm.

At a check node extrinsic LLRs are computed by decoding the associated single parity-check component code. The message received by  a symbol node from its $j$th neighboring check node,  $j=1,\dots,J$, can be written as
\begin{equation}\label{eq:checkupdate}
\beta^{(j)}=2 \mathrm{arctanh} \left(\prod_{k' \neq k} \tanh(z^{(k')}/2) \right) \enspace ,
\end{equation}
where $z^{(k')}$, $k'=\{1,\dots, K\} \setminus k$,  are the messages that this check node has received from its other adjacent symbol nodes.
The incoming extrinsic LLRs are then combined with the intrinsic channel LLR $\alpha$ of the considered symbol to give the LLRs
\begin{equation}\label{eq:variableupdate}
z^{(j)}=\alpha+\sum_{j' \neq j} \beta^{(j')} \enspace, \quad j=1,\dots,J \enspace ,
\end{equation}
which form the messages to be sent back to the check nodes.
Initially, before the first decoding iteration, the LLRs are set to $z^{(j)}=\alpha$ for symbols at times $t=1,\dots,L+J$. For all other $t$ the code symbols are defined to be zero, which implies that $z^{(j)}=\infty$ through all iterations. This initial condition automatically takes into account the lower check node degrees at the beginning and the end of the Tanner graph. 

We consider the standard parallel updating schedule where, in each decoding iteration, first all check nodes and then all symbol nodes are updated according to (\ref{eq:checkupdate}) and (\ref{eq:variableupdate}), respectively.
The messages computed in this way are true LLRs as long as they are produced from independent observations.  
The following theorem guarantees that the number of independent iterations possible on the Tanner graph of the block code, produced by terminating convolutional codes from the ensemble $\mathcal {C}_P(J,2J,M)$, can be made arbitrarily large.
\begin{theorem} For any length $L$ there exists   
a code in $\mathcal {C}_P(J,2J,M)$ for which the number of independent decoding iterations, $l_0$, satisfies 
\[
l_0> \frac{\log M}{2\log (2J-1)(J-1)}- c_1 \enspace ,
\] 
where the constant $c_1$ does not depend on $M$. \hfill $\square$
\end{theorem}
The proof of this theorem is based on an analogous theorem for LDPC block codes given in~\cite{rel_comm_micha}. 
Given that all messages are formed from independent observations, it is possible to calculate the evolution of their exact probability density functions (pdfs) during the iterations 
\cite{RSU:irrLDPC} (density evolution).
These pdfs can be used to find an upper bound on the smallest channel SNR (convergence threshold) for which the error probability converges to zero as the number of iterations goes to infinity.
Since, in general,  density evolution must be performed numerically, we follow the approach in \cite{rel_comm_micha} and estimate the asymptotic convergence rate by observing aside from the pdfs  of the LLRs $z^{(j)}$ also their Bhattacharyya parameter.

For regular LDPC block codes, the distribution of the messages exchanged in iteration $\ell$ is the same for all nodes regardless of their position within the graph.
Likewise, for the random irregular code ensembles considered in \cite{RSU:irrLDPC}, the message distributions are averaged over all codes and only a single mixture density 
need be 
considered for all check nodes and all symbol nodes, respectively.
Looking at the flow of messages in the Tanner graph it can be seen that this is not true in our case.
\begin{figure}
\centering
\begin{psfrags} \small
\psfrag{t}[c][c]{$t$}
\psfrag{channel}[c][c]{channel}
\psfrag{t-1}[c][c]{$t-1$}
\psfrag{t-2}[c][c]{$t-2$}
\psfrag{t+1}[c][c]{$t+1$}
\psfrag{t+2}[c][c]{$t+2$}
\psfrag{(a)}[c][c]{(a)} \psfrag{(b)}[c][c]{(b)}
\includegraphics[width=0.9\linewidth]{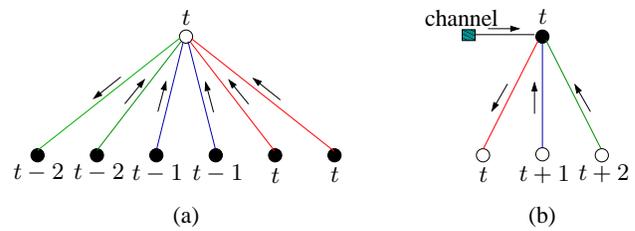}
\caption{Illustration of the messages (a) to a symbol node and (b) to a check node for the case $J=3$.} 
\label{fig:messages}
\end{psfrags}
\end{figure}
\begin{figure}[t] \centering
\begin{psfrags}\scriptsize
\psfrag{time=1}[c][c]{$t=1$}
\psfrag{time=2}[c][c]{$t=2$}
\psfrag{time=3}[c][c]{$t=3$}
\psfrag{time=4}[c][c]{$t=4$}
\psfrag{time=11}[c][c]{$t=1$}
\psfrag{time=21}[c][c]{$t=2$}
\psfrag{time=22}[c][c]{$t=2$}
\psfrag{time=23}[c][c]{$t=2$}
\psfrag{time=31}[c][c]{$t=3$}
{
{\includegraphics[width=\linewidth]{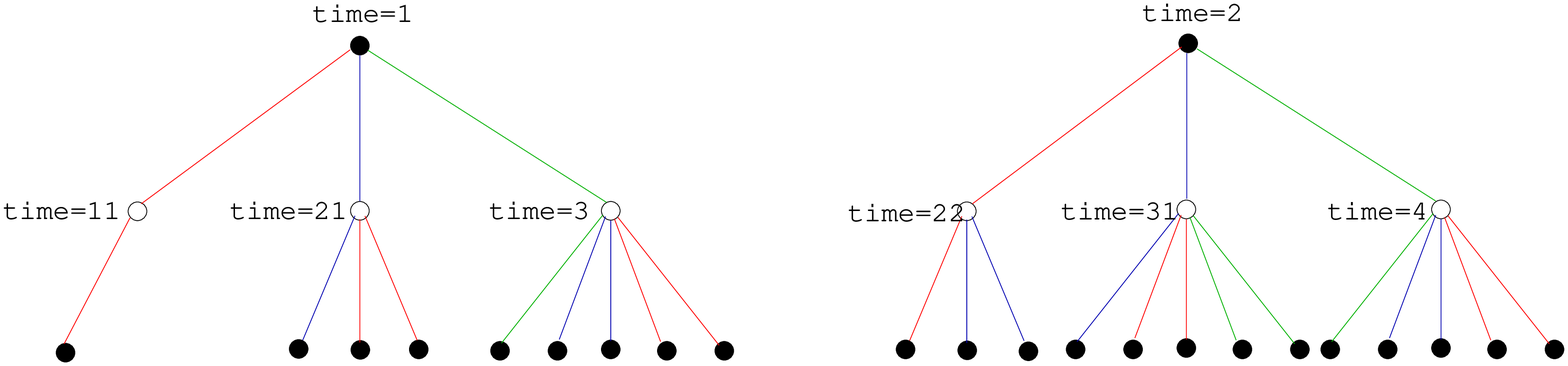}
}}

\vspace{8mm}
\scalebox{0.9}{
\psfrag{time=3}[c][c]{$t=3$}
\psfrag{time=4}[c][c]{$t=4$}
\psfrag{time=5}[c][c]{$t=5$}
\psfrag{time=31}[c][c]{$t=3$}
{
{\includegraphics[width=0.7\linewidth]{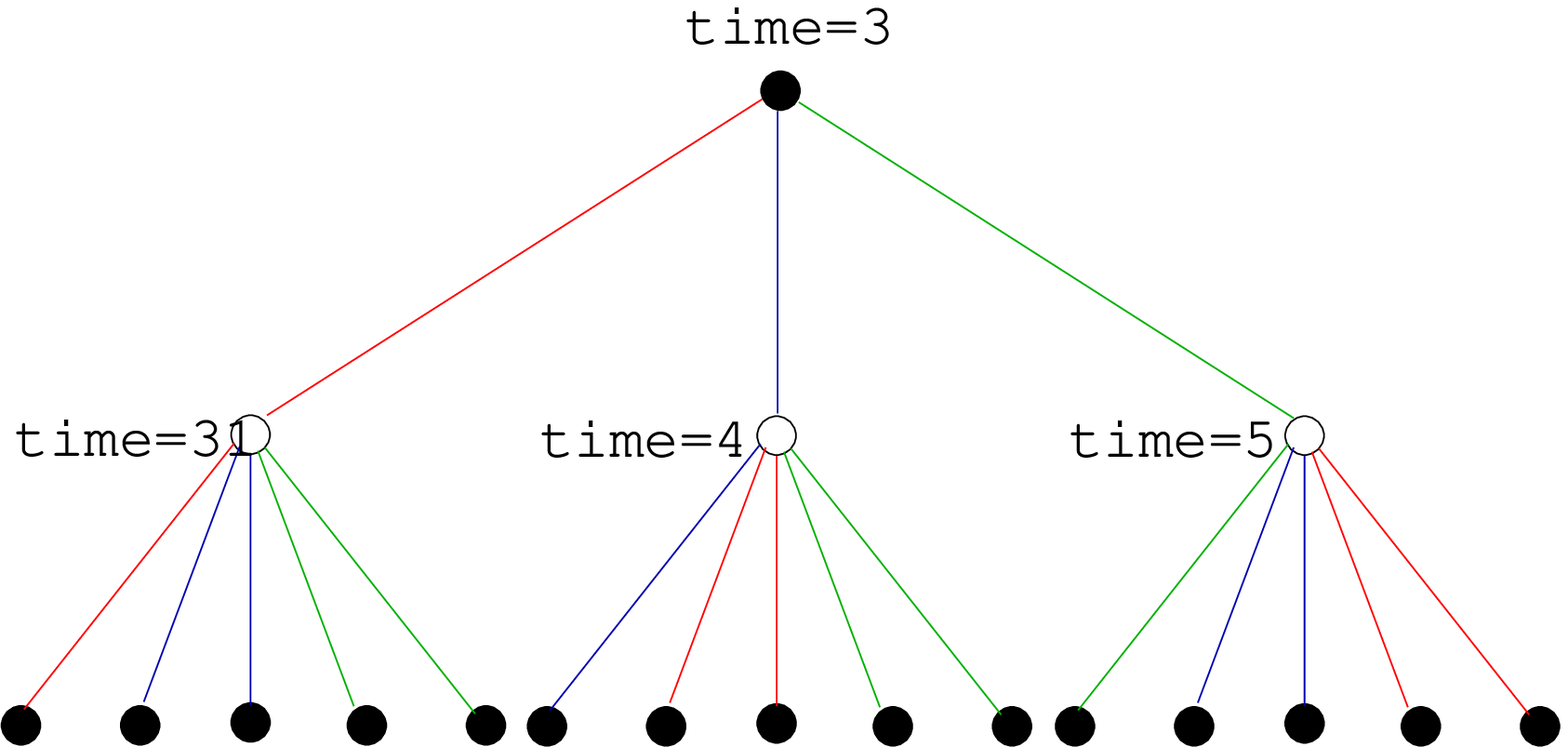}
}}}
\end{psfrags}
\caption{The first level of the computation trees for $t=1,2,3$ with $J=3$.}
\label{fig:comp_tree}
\vspace*{-1.5em}
\end{figure}
As shown in Fig.~\ref{fig:messages}(a), messages $z^{(k')}$  in (\ref{eq:checkupdate}) come from  nodes belonging to different time instants.
The same holds for the messages $\beta^{(j')}$ that are combined in (\ref{eq:variableupdate}) (see Figure \ref{fig:messages}(b)).
Fig.~\ref{fig:comp_tree} shows the first level of the corresponding decoding computation trees for the first three symbol levels in the case $J=3$.
Although only the first $J-1$ levels of check nodes have lower degrees, and hence provide better protection, their effect 
propagates through the complete graph.
Consequently, while nodes at the same time instant behave identically, the messages from nodes at different times behave differently and must all be tracked separately.
To take this into account, for each level different pdfs  
must be computed in every iteration for both $\beta^{(j)}$ and $z^{(j)}$, $j=1,\dots,J$. 

Consider now the $\ell$th decoding iteration, where $1\leq \ell \leq \ell_0$. 
Let $\varphi_{t,t+k}^{(\ell)}(z|0)$ and $\varphi_{t,t+k}^{(\ell)}(z|1)$ be the pdfs of messages  sent from the node of 
a code symbol $v_m$ at time $t$ to one of its neighboring check nodes at time $t+k$, conditioned on $v_m=0$ and $v_m=1$, respectively.
The Bhattacharyya parameter $B^{(\ell)}_{t,t+k}$ of these messages, $k=0,\dots,J-1$, is equal to 
\begin{equation}\label{eq:bhat}
B_{t,t+k}^{(\ell)}=\int_{-\infty}^{\infty} \sqrt{\varphi_{t,t+k}^{(\ell)}(z|0)\varphi_{t,t+k}^{(\ell)}(z|1)} dz \enspace .
\end{equation}
For the intrinsic channel LLRs $\alpha$, the Bhattacharyya parameter is derived analogously from the channel transition pdf and is denoted as $A$.
The following lemma, analogous to Lemma~1 in \cite{rel_comm_micha}, connects the Bhattacharyya parameters corresponding to the LLRs of two consecutive decoding iterations.
\begin{lemma}
The Bhattacharyya parameter $B_{t,t+k}^{(\ell)}$, defined by (\ref{eq:bhat}) for $\ell=1,\dots,\ell_0$ and $k=0,\dots,J-1$, satisfies the following inequality
\begin{equation}\label{eq:lem1}
B^{(\ell)}_{t,t+k} < A \prod_{k' \neq k} \left( B^{(\ell-1)}_{t,t+k'} + \sum_{i' \neq k'} (B^{(\ell-1)}_{t+k'-i',t+k'})^2\right) \enspace ,   
\end{equation}
where $k',i' \in \{0,\dots,J-1\}$ and $B^{(0)}_{t,t+k}=0$. \hfill $\square$
\end{lemma} 
If we define $B\stxt{max}^{(\ell)}$ as the largest value of $B_{t,t+k}^{(\ell)}$ over all edges in the graph, i.e.,
\begin{equation}\label{eq:bmax}
B\stxt{max}^{(\ell)}=\max_{t,k} B_{t,t+k}^{(\ell)} \enspace, \quad k \in \{0,\dots,J-1\} \enspace ,
\end{equation}
then it follows from (\ref{eq:lem1}) that
\begin{equation}\label{eq:bhatmax}
B^{(\ell)}_{t,t+k} \leq B\stxt{max}^{(\ell)} < A\left((2J-1)B\stxt{max}^{(\ell-1)}\right)^{J-1} \enspace.
\end{equation}
Suppose now that after some iteration $\ell'<\ell_0$ all Bhattacharyya parameters $B^{(\ell')}_{t,t+k}$ are smaller than the {\em breakout value}
\begin{equation}\label{eq:breakout}
B\stxt{br}=A^{-1/(J-1)}(2J-1)^{-(J-1)/(J-2)} \enspace .
\end{equation}
As described in \cite{rel_comm_micha},  it follows then from (\ref{eq:bhatmax}) that after $\ell_0$ decoding iterations the bit error probability $P\stxt{b}(t)$ for symbols at an arbitrary time $t$  satisfies
\[
P\stxt{b}(t) < B\stxt{max}^{(\ell_0)} < \left( \frac{B\stxt{max}^{(\ell')}}{B\stxt{br}}\right)^{(J-1)^{\ell_0-\ell'}} \enspace.
\]
This shows that the bit error probability of all code symbols converges to zero at least double exponentially with the number of decoding iterations.

To determine convergence thresholds for terminated codes from the ensemble $\mathcal {C}_P(J,2J,M)$ it is possible to numerically evaluate, iteration by iteration, the different pdfs $\varphi_{t,t+k}^{(\ell)}(z|\cdot)$  for all time instants $t$.
For proving that the bit error probabilities of all symbols converge to zero, it is sufficient to check that $B\stxt{max}^{(\ell')}$ is below the  breakout value $B\stxt{br}$ after some number of iterations $\ell'$.
The convergence threshold for an ensemble of codes  can be found by testing this condition for different channel values.

Note that, in addition to the node degrees $J$ and $K$, the value $L$ is another parameter that influences the result.
For small $L$ there is a significant rate loss due to the termination, and results for the erasure channel show  that the threshold can even surpass the capacity of codes with rate $R=1/2$ \cite{allerton}, which can be explained by the large fraction of strong check nodes of low degree.
It has also been observed in \cite{allerton} that for large $L$ the threshold for terminated convolutional codes remains constant.
This is especially interesting since with increasing $L$ the  degree distribution becomes closer and closer to that of a regular $(J,2J)$ block code, which has a significantly weaker threshold.
Hence, the improved threshold can be attributed to the special structure of the Tanner graph imposed by the convolutional nature of the codes and not only to the ratio 
of stronger to weaker nodes.

While increasing $L$ reduces the rate loss, the computational burden of performing density evolution becomes increasingly difficult for larger $L$.
Both the number of different pdfs to be tracked 
and the number of iterations until the effect from strong nodes at the ends of the graph carries through to the levels in the middle increase with $L$.
Also, for the AWGN channel, the complexity of density evolution is much higher than for the erasure channel, where a simple one-dimensional recursion formula can be used. 
In the next section we consider therefore a sliding window updating schedule that reduces the number of operations required for the threshold computation.

\section{Threshold Computation: A Sliding Window Approach}
In the standard parallel updating schedule, considered in the previous section, first all check nodes and then all symbol nodes are activated in each iteration. 
This is convenient for an analysis of decoding since then the computation trees of all symbols have a very regular structure.
An alternative schedule, where a symbol node is activated whenever a message is demanded by a neighboring check node, was considered in \cite{ReducedCompl}. The check nodes are activated one by one. It has been observed in computer simulations that such an on-demand symbol node update can reduce the required number of decoding iterations.

In general, any arbitrary node activation order in the decoding will result in a particular shape of the computation trees of  the different code symbols.
Since, for any finite number of node activations, the depths of the computation trees will be finite as well, it follows that these trees  can be covered by trees corresponding to a parallel updating schedule with a sufficient number of iterations.
Consequently, any node updating schedule can be interpreted as a parallel schedule where certain node activations are omitted.
But, under the independence assumption, such an omission of additional side information can never improve the performance of decoding, which results in the following proposition. (A similar result has also been obtained in~\cite{and_thesis}.) 
\begin{prop}\label{lem:schedule}
Consider density evolution with an arbitrary node activation order. Assume that the breakout value condition, described in the previous section, is satisfied after a specific number of node activations.
Then this condition will also be satisfied for a standard parallel updating schedule after a sufficiently large number of iterations. \hfill $\square$
\end{prop}

Let us now consider the following updating schedule in density evolution.
In a window from level $t=t'$ to level $t=\min(t'+W-1,L/2)$, $W\leq L/2$, all symbol nodes are activated one by one according to (\ref{eq:variableupdate}). Check nodes are activated according to (\ref{eq:checkupdate}) whenever a neighboring symbol node demands a message.
The starting position of this updating window of size $W$ is initialized by $t'=1$.
The nodes within the window are updated repeatedly until the error probability at level $t'$ reaches some value that corresponds to a sufficiently small  Bhattacharyya parameter $B_0$, $0< B_0 < B\stxt{br}$.
Then the window is shifted by increasing $t'$ by one.
(Note that we only have to consider levels $t=1,\dots,L/2$ because of the symmetry within the Tanner graph.)

\begin{figure} \centering
\scalebox{0.8}{
\begin{psfrags} \small
\psfrag{updates per level}[c][c]{updates per level}
\psfrag{level}{t}
\psfrag{updates (iterations)}{updates}
\psfrag{Pb}{$P\stxt{b}$}  \psfrag{t=1 }{$t=1$}\psfrag{t=5 }{$t=5$}\psfrag{t=10 }{$t=10$}
\psfrag{t=15 }{$t=15$} \psfrag{t=20 }{\hspace*{-0.8ex}$t=20$} \psfrag{t=25 and 30 }{$t=25$ and $30$}
\psfrag{2M}[c][c]{$2M$}
\psfrag{L}[c][c]{$L+2$}
{\includegraphics[width=1.25\linewidth]{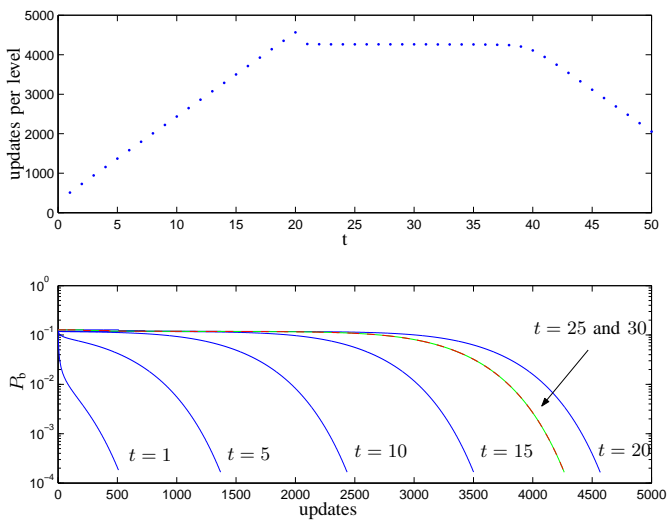}}
\end{psfrags}}
\caption{Updates per level and bit error probability behavior for $J=3$, $L=100$, and $W=20$ at $E\stxt{b}/N_0=0.55$ dB.}
\label{fig:exawgn}
\vspace*{-1.5em}
\end{figure}
For $J=3$, Fig.~\ref{fig:exawgn} shows the required number of updates per level (computational complexity) and the bit error probability at different levels as a function of the updates. In this example the signal-to-noise ratio $E\stxt{b}/N_0=0.55$ dB of the AWGN channel  corresponds to the estimated threshold, which explains the high number of required updates.
Due to the window approach the number of updates per level increases until it reaches its maximum at $t=20$. After that it remains at a constant level until the window reaches the middle of the Tanner graph, where (symmetry) effects from the other end of the graph result in a reduction of updates.
It has been observed that after the window size $W$ exceeds a certain minimal value, the bit error curves no longer change if  $W$ is further increased.  This helps us to choose the parameter $W$ in our calculations. 
But more importantly, we can conclude that this approach is 
as good as a non-windowed updating schedule.

It can also be seen in Fig.~\ref{fig:exawgn} that the bit error curves for levels $t=25$ and $t=30$ are almost indistinguishable. This is actually the case  in the complete region of levels where the number of required updates stays constant.
This indicates that the performed calculations tend to repeat themselves at different window positions.
From this we may conclude that the effect of the strong nodes at the ends of the Tanner graph carries through to the middle independently of the termination length $L$.
This confirms the observation on the erasure channel in \cite{allerton} that the threshold remains constant for large $L$ and that the number of iterations required for reaching a certain bit error probability at a node in the middle of the graph  increases linearly with $L$.
Furthermore, it suggests that detecting convergence at the first levels is sufficient to determine the overall convergence threshold.

The same observations can be made for the erasure channel, for which we 
can state the following result.

\begin{prop}\label{th:allL}
Consider density evolution on the erasure channel with the window updating schedule described above for an arbitrary termination length $L$. Starting from $t'=1$, assume that $t'$ is increased as soon as the Bhattacharyya parameter at level $t'$ is below some value $B_0<B\stxt{br}$. Under these conditions, if the window can be  shifted at least $J$ times, then $B_0$ can  be reached at all $t$, $1 \leq t \leq L$. \hfill $\square$
\end{prop}

This proposition can be proved by induction when the updating window is initialized by the same pdfs at  different window positions $t'$. Here we 
make use of the fact that the pdfs $\varphi_{t,t+k}^{(\ell)}(z|\cdot)$ computed within density evolution can be ordered in terms of quality. For the erasure channel this follows from the fact that the pdfs are described by a single parameter, the probability of erasure.
For other channels such an ordering of the pdfs $\varphi_{t,t+k}^{(\ell)}(z|\cdot)$ is not obvious.
However, we conjecture that the statement of Proposition~\ref{th:allL} is  true for arbitrary binary-input memoryless channels.

In Table \ref{tab:thres} the estimated thresholds $(E\stxt{b}/N_0)^{*}$ for the AWGN channel are presented for different $J$. In the computations the  values of $L$ were chosen such that there is a rate loss of $2\%$, i.e., $R=0.49$. 
Assuming that the thresholds are independent of the termination length $L$, the right hand side of the table shows the corresponding threshold values $(E\stxt{b}/N_0)^{**}$ for the case $L\rightarrow \infty$. 
Similar to the erasure channel results, the thresholds are much better than those of the corresponding regular LDPC block codes (e.g., 1.11 dB for $J=3$), and 
tend to the capacity limit of rate $R=1/2$ codes with increasing $J$. 
\begin{table}
\begin{center}                                
\begin{tabular}{|c||c|r||c|r|}
\hline
$(J,K)$&$R$&$(E\stxt{b}/N_0)^{*}$& $R$&$(E\stxt{b}/N_0)^{**}$\\
\hline
(3,6)&0.49&0.55 dB& 0.5 & 0.46 dB \\
(4,8)&0.49&0.35 dB& 0.5 & 0.26 dB\\
(5,10)&0.49&0.30 dB& 0.5 & 0.21 dB\\
\hline
\end{tabular}
\end{center}
\caption{Thresholds for the ensembles $\mathcal {C}_P(J,2J,M)$ with 
different $J$.}
\label{tab:thres}
\vspace*{-2.5em}
\end{table}

\vspace*{-0.65em}
\section{Acknowledgments}
This research was supported in part by NSF Grant CCR-02-05310, NASA Grant NAG5-12792,  the State of Indiana 21st Century Science and Technology Fund, and the German research council Deutsche Forschungsgemeinschaft under Grant Bo 867/12.

\vspace*{-1.2em}
\bibliographystyle{ieeetr}
\bibliography{paper_bib}
\end{document}